\documentclass[aps,showpacs,twocolumn,superscriptaddress,prl]{revtex4-1}
\usepackage{graphicx,color,epsfig,rotate}
\usepackage{color}
\usepackage{graphicx}
\usepackage{dcolumn}
\usepackage{amssymb}
\usepackage{amsmath}
\usepackage{amsfonts}
\usepackage{docs}%
\usepackage{bm}%
\usepackage{wrapfig}
\usepackage[colorlinks=true,linkcolor=blue,pagecolor=blue,filecolor=blue,menucolor=blue,urlcolor=blue,citecolor=blue,anchorcolor=blue]{hyperref}%
\usepackage{bm}
\usepackage{graphics}
\usepackage{graphicx}
\usepackage{color}
\usepackage{amssymb}
\usepackage{standalone}
\usepackage[toc,page]{appendix}
\usepackage{import}
\usepackage{enumerate}
\usepackage{sidecap}

\newcommand{\bnabla}{{\boldsymbol \nabla}}

\usepackage{amssymb,amsmath,bm}

\begin{document}

\title{Pure Odd Frequency Superconductivity at the Cores of Proximity Vortices}

\author{Mohammad Alidoust}
\affiliation{Department of Physics, Faculty of Sciences, University of Isfahan, Hezar Jerib Avenue, Isfahan 81746-73441, Iran}
\author{Alexander Zyuzin}
\affiliation{Department of Theoretical Physics, The Royal Institute of Technology, Stockholm, SE-10691 Sweden}
\affiliation{A. F. Ioffe Physical - Technical Institute, 194021 St. Petersburg, Russia}
\author{Klaus Halterman}
\affiliation{Michelson Lab, Physics Division, Naval Air Warfare Center, China Lake, California 93555, USA}

\date{\today}

\begin{abstract}
After more than a decade, direct observation of the odd frequency triplet pairing
state in 
superconducting hybrid structures remains elusive. 
We propose an
experimentally feasible setup that
can 
unambiguously reveal the zero energy peak due to proximity-induced 
\textit{equal spin} superconducting triplet correlations. 
We theoretically investigate
a two dimensional Josephson junction in the \textit{diffusive} regime.
The nanostructure consists of
 a normal metal sandwiched between
 two ferromagnetic layers with spiral magnetization patterns.
 By applying an external magnetic field perpendicular to the junction plane,  
 vortices nucleate in the normal metal. 
 The calculated energy and spatially resolved density of states, along
 with the pair potential,
 reveal that remarkably, only triplet Cooper pairs survive in the vortex cores.
 These isolated odd frequency triplet correlations result in well defined
 zero energy peaks in the local density of states that can be identified 
 through tunneling spectroscopy experiments.
Moreover, the diffusive regime considered here 
rules out the possibility of Andreev bound states
in the vortex core as contributors to the zero energy peaks.
 \end{abstract}


\maketitle

{\it Introduction.}~
In analogy to the pairing mechanism in $^3${\rm He}, spin triplet Cooper pairing was predicted 
to coexist with spin singlet correlations in hybrid structures consisting of
 $s$-wave superconductors (S) and inhomogeneous ferromagnets (IFMs)\cite{Buzdin2005,first,Bakurskiy1,Keizer2006,Halterman2007,Bobkova,Khaydukov,Moor,singh,Eschrig2015,longrg,jason_natphys}. 
For these types of systems, 
spin triplet correlations with nonzero ($\pm 1$) projections
 along a given spin quantization 
axis can
result in long range proximity effects \cite{Buzdin2005,first}. 
It was argued that 
traces of the triplet pairing state could be 
revealed in measurements of the
critical supercurrent \cite{Buzdin2005,first,kup_rmp,crnt_1,crnt_2,crnt_hoz,longrg,Eschrig2015,Khaydukov,Keizer2006,Baker} and local density of states (LDOS) \cite{buz_zep,golubov,klaus_zep,golu}. In the former case, 
the critical supercurrent should show 
a slow damping behavior as a function of 
spin singlet depairing factors (such as the thickness of a uniform
magnetic layer), 
while in the later case,  
the LDOS should exhibit a peak at zero energy. 
Unfortunately, an unambiguous  and 
direct observation of the spin triplet pairing
state in 
F/S hybrid platforms 
remains elusive due to the 
difficulty in  isolating the triplet pairs entirely, 
even when a
half-metallic ferromagnet is incorporated
  \cite{klaus_zep,bernard,km_hf,singh,buzhf,Eschrig2015}. 
Thus, it is preferable to find a practical way to manipulate the
pair correlations so that
 the singlet and triplet components occupy separate regions of space.
In contrast to current approaches~\cite{buzhf,crnt_2,golu,crnt_1, golubov, Buzdin2005,first,klaus_zep,bernard,km_hf,singh,Eschrig2015,longrg,jason_natphys,golu},
controlling the pair correlations in this way
can be achieved by
applying a magnetic field to the F/S structure \cite{longrg},
inducing proximity vortices with normal state cores. 
This may consequently create a favorable situation where  the
singlet and triplet pair correlations can 
be fully separated at the vortex cores.

The first experimental 
observation of nonmagnetic 
proximity induced vortices recently
occurred in 
two dimensional  normal 
metal (N) Josephson junctions \cite{vrtx_exp}. 
It was observed that
applying an external magnetic field perpendicular to a wide 
SNS Josephson junction causes 
nucleation of a
vortex lattice in the normal metal parallel to the SN interfaces. 
The 
number of induced vortices depends on the intensity of the
externally applied magnetic field.
The proximity-induced vortices in two dimensional 
Josephson structures was  
first
discussed theoretically  
in connection with the Fraunhofer and anomalous critical supercurrent
 responses in 
 Josephson junctions with both 
 normal metal \cite{berg_fr,mj_nf} and ferromagnetic elements \cite{longrg}. 
 This concept was also recently extended to disordered surface states of topological insulators 
 and Dirac materials in the
 quasiclassical regime
 \cite{ZAD_3DTI}. 

\begin{figure}[t!]
\includegraphics[width=7.10cm,height=5.0cm]{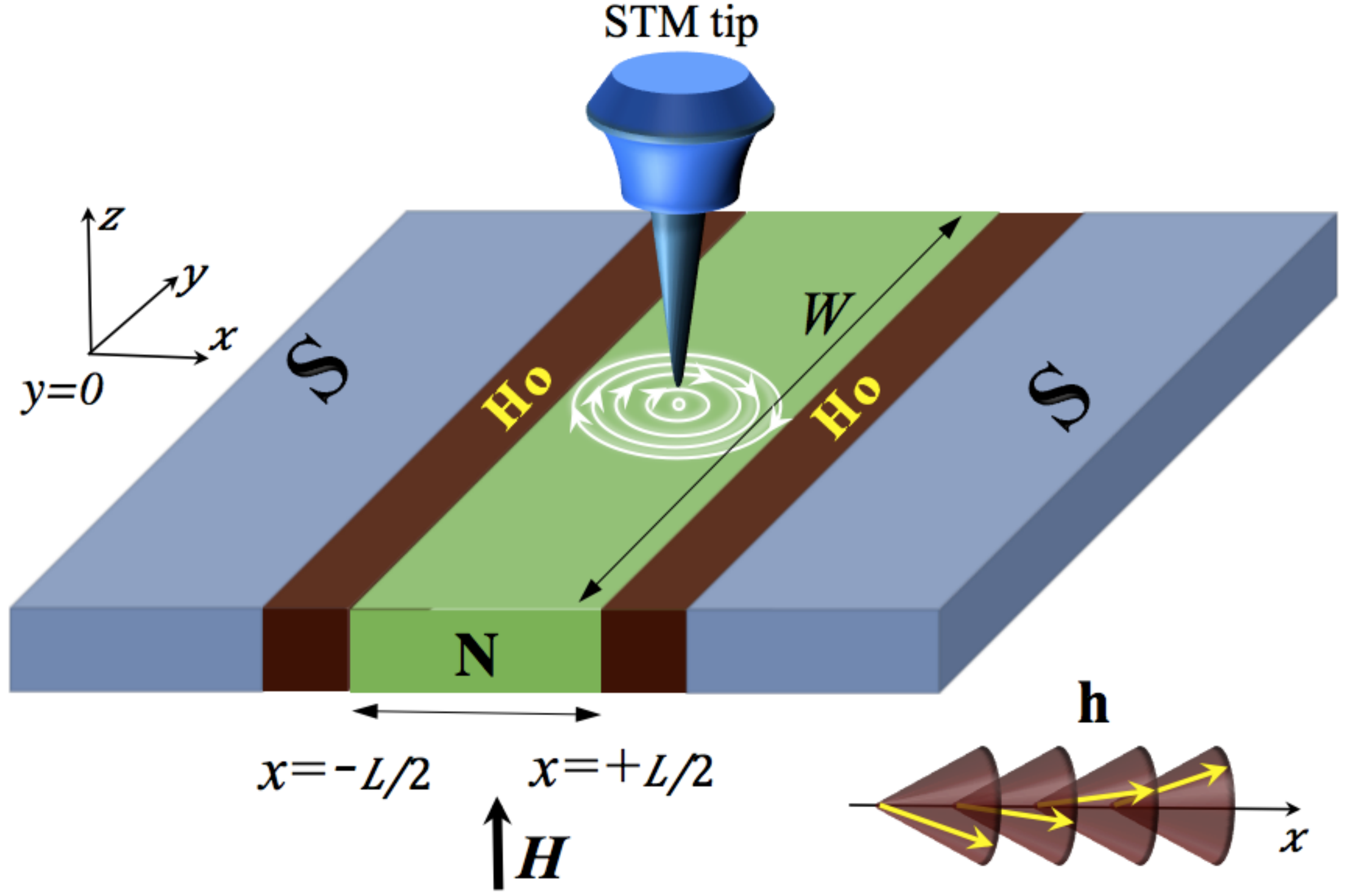}
\caption{\label{fig1} (Color online) 
Schematic of the S-Ho/N/Ho-S junction subject to a perpendicular external magnetic field ${\bf H}$. The junction plane resides in the $z=0$ plane and the N/Ho interfaces 
are located at $x=\pm L/2$. 
The junction has a length and width of $L$ and $W$, respectively. The two helimagnets (Holmium type) with internal fields ${\bf h}$ (see text) are attached to the diffusive normal metal (N) solely for producing spin-1 triplet pair correlations. The perpendicular external magnetic field induces proximity vortices in the N region depicted schematically.
 }
\end{figure}

In this paper, we study the 
{\it diffusive} S-${\rm Ho}$/N/${\rm Ho}$-S Josephson junction structure 
shown in Fig.~\ref{fig1} 
as a system for fully isolating the odd frequency 
spin-1 superconducting triplet correlations.
The existence of the triplet pairs 
is directly revealed in the
form of DOS signatures. 
The role of the Holmium (${\rm Ho}$) layers is
that of a
spin-1 triplet pairing source, while
the superconducting phase gradient across the junction drives the
triplet pairs into the N region. 
By taking advantage of the fact that an external magnetic field  
applied perpendicularly to
the junction plane induces vortices in the N region, while
expelling the spin singlet pairs from the vortex centers (creating a normal core), 
we demonstrate that spin-1 triplet correlations occupy the  normal core region, 
as 
revealed through peaks in the zero energy DOS. We support our findings by a
 spin parameterization technique to the Green function of system that allows for fully
 identifying the behavior of each individual pair correlation \cite{longrg}.
Since the N layer is a diffusive metal with  
numerous strong scattering sources, the superconducting coherence
length is much larger than 
the mean free path, and therefore bound states  cannot
form at the centers of the
vortices, ruling out Andreev bound states as contributors to the 
zero energy peak (ZEP). Consequently, 
the spin-1 triplet channel is highly dominant within the vortex core,
causing the ZEP in the DOS.

{\it Results and Discussions.}~
It is now firmly established
that the electronic properties of a diffusive hybrid superconducting structure can be described by 
the Usadel equation within the quasiclassical framework \cite{Buzdin2005,first,kup_rmp}. 
The Usadel equation in the normal region reads \cite{usadel}: 
\begin{eqnarray} \label{eq:usdl}
D \hat{\bnabla}(\hat{G} \hat{\bnabla} \hat{G}) +i  [\varepsilon\hat{\rho}_z, \hat{G}]=0,
~~\hat{G}(\varepsilon,\mathbf{ R})=\left( \begin{array}{cc}
G^A& G^K \\
0 & G^R
\end{array} \right),~~~~~
\end{eqnarray}
where $D$ represents the diffusion constant in the N and S regions and 
$\varepsilon$  is  the quasiparticle energy measured from the Fermi level. We normalize all lengths by the superconducting coherence length, $\xi_S=\sqrt{\hbar D/|\Delta_0|}$, energies by the superconducting gap at zero temperature, $|\Delta_0|$, and adopt  natural units where $\hbar=k_B=1$. 
The Green function $\hat{G}(\varepsilon,\mathbf{ R})$ is composed of the advanced, 
$G^A(\varepsilon,\mathbf{ R})$, 
retarded, $G^R(\varepsilon,\mathbf{ R})$, and Keldysh, $G^K(\varepsilon,\mathbf{ R})$, propagators, which carry 
the complete  
physical information of the system considered. 
In the presence of an external magnetic field, ${\bf H}=(0,0,H_z)$, directed 
perpendicular to the junction plane, 
the derivatives can be replaced by their covariants, i.e. $\hat{\bnabla} = \bnabla - [ie{\bm A}\hat{\rho}_z, ...].$ 
Here ${\bm A}$ is the vector potential associated with the external field ${\bm H}$. 
In equilibrium, as considered throughout the paper, the advanced and Keldysh propagators 
can be expressed via the Retarded Green function.
 In this case, one can show that
 $G^A(\varepsilon,\mathbf{ R})=-\{\hat{\rho}_zG^R(\varepsilon,\mathbf{ R}) \hat{\rho}_z\}^\dag$ and $G^K(\varepsilon,\mathbf{ R})=\{G^R(\varepsilon,\mathbf{ R})-G^A(\varepsilon,\mathbf{ R})\}\tanh(\varepsilon k_BT/2)$,
 where $k_B$ is the Boltzmann constant,  and the system temperature is denoted by $T$. 
 Therefore, it suffices to focus on the retarded Green function, and then eventually construct
  the total propagator using  the  simple relations above. 
  One useful limit for F/S 
  structures is the so-called low proximity limit. 
  This limit permits  linearization of the Green function, yielding
 a  linear set of  differential equations that are in general coupled \cite{ZAD_3DTI,longrg}. 
 Although highly useful transport 
 characteristics 
 can be captured in this limit, 
 the full proximity regime allows 
 for the study of energy-resolved and spatially-resolved 
 DOS, and 
other
relevant physical quantities. 
Hence,
we first employ the 
full proximity 
limit, 
resulting in a complex set of
nonlinear coupled differential equations \cite{longrg} and then
compliment  our findings with a spin parameterization technique in the low 
proximity limit.

In establishing  a numerically stable algorithm in the \textit{full} proximity limit, 
we use 
the so-called Riccati parametrization \cite{riccati}, where it is convenient to
introduce two correlated 
functions $\gamma$ and $\tilde{\gamma}$, which are in effect unknown 2$\times$2  matrices. 
In this parameterization scheme, the retarded Green function takes the following form:          
\begin{eqnarray} \label{eq:gf_riccati}
G^R(\varepsilon,\mathbf{ R})=\left( \begin{array}{cc}
(1-\gamma\tilde{\gamma})\Gamma& 2\gamma\tilde{\Gamma} \\
2\tilde{\gamma}\Gamma & (\tilde{\gamma}\gamma-1)\tilde{\Gamma}
\end{array} \right), 
\end{eqnarray}
in which $\Gamma=(1+\gamma\tilde{\gamma})^{-1}$ and $ \tilde{\Gamma}=(1+\tilde{\gamma}\gamma)^{-1}$. 
Substituting the Riccati parameterized Green function into the Usadel equation, Eq.~(\ref{eq:usdl}), and considering the external magnetic field, we arrive at the following equations 
for $\gamma$ and $\tilde{\gamma}$ in the N region of Fig. \ref{fig1}:
\begin{subequations}\label{eq:usdl_riccati}
\begin{eqnarray}\nonumber
&&\partial_{k,k'}^2\gamma-2(\partial_{k,k'}\gamma)\tilde{\gamma}\Gamma\partial_{k,k'}\gamma -(2eH_zk')^2\left\{2\Gamma-1\right\}\gamma\nonumber\\&&-4ieH_zk'\left\{ \partial_{k}\gamma-(\partial_{k}\gamma)\tilde{\Gamma}-\Gamma\partial_{k}\gamma\right\}=-2i\frac{\varepsilon}{D}\gamma, \label{eq:usdl_riccati1}\\&&
\partial_{k,k'}^2\tilde{\gamma}-2(\partial_{k,k'}\tilde{\gamma})\gamma\tilde{\Gamma}\partial_{k,k'}\tilde{\gamma}-(2eH_zk')^2\left\{2\tilde{\Gamma}-1\right\}\tilde{\gamma}\nonumber\\&&+4ieH_zk'\left\{ \partial_{k}\tilde{\gamma}-(\partial_{k}\tilde{\gamma})\Gamma-\tilde{\Gamma}\partial_{k}\tilde{\gamma}\right\}=-2i\frac{\varepsilon}{D}\tilde{\gamma}. \label{eq:usdl_riccati2}
\end{eqnarray}
\end{subequations}
For compactness, we have defined  
$k\equiv x$ and $k'\equiv y$ for the spatial coordinates so that $\partial_{k,k'}\equiv \partial_x+\partial_y$.
We have also employed the Coulomb gauge, so that ${\bm \nabla}\cdot{\bm A}=0$. 

We consider a realistic situation 
where the junctions are well described by a 
tunneling process \cite{boundary_c}. The appropriate boundary conditions
for this regime are the Kupriyanov-Zaitsev boundary 
conditions \cite{boundary_c}:
\begin{eqnarray}\label{eq:bc}
\zeta \hat{G} \mathbf{ n}\cdot\hat{\bnabla} \hat{G} = [\hat{G}, \hat{G}_{\text{S}}],~~ G^R_{\text{S}}=\left( \begin{array}{cc}
{\cal C}& {\cal S} e^{+i\varphi}\\
{\cal S}e^{-i\varphi} & -{\cal C}
\end{array} \right),~~
\end{eqnarray}
where $\zeta$ is the ratio of the
barrier resistance  to the resistivity of the normal layer, 
and the components of the 
retarded superconducting bulk solution \cite{boundary_c}, 
can be expressed by ${\cal C}\equiv \cosh\theta\sigma_0$ and ${\cal S}\equiv i\sinh\theta\sigma_y$, in which $\theta=\text{atanh}(\Delta/\varepsilon)$. The superconducting phase is denoted by $\varphi$ and 
the unit vector normal to the interfaces is denoted
 by $\mathbf{ n}$. Inserting the Riccati parameterized Green function into the 
 boundary conditions, Eq.~(\ref{eq:bc}), we find the following first order differential equations
at $x=\mp L/2$ with $\varphi=\pm\phi/2$:
\begin{subequations}\label{eq:bc_riccati}
\begin{eqnarray}
&&\partial_{k}\gamma +2ieH_zk' \gamma=\pm(2\frac{{\cal C}}{{\cal S}}+\gamma e^{\mp i\phi/2}-\frac{e^{\pm i\phi/2}}{\gamma})\frac{{\cal S}\gamma}{\zeta},~~~\\
&&\partial_{k}\tilde{\gamma} -2ieH_zk' \tilde{\gamma}=\pm(2\frac{{\cal C}}{{\cal S}}+\tilde{\gamma} e^{\pm i\phi/2}-\frac{e^{\mp i\phi/2}}{\tilde{\gamma}}) \frac{{\cal S}\tilde{\gamma}}{\zeta}.
\end{eqnarray}
\end{subequations}
Next, to generate spin-1 triplet correlations and have them 
occupy the N region, several practical ways can be considered \cite{Eschrig2015}. 
For example, the triplets can be generated in a SF/N/FS
type  junction 
with the aid of uniform noncollinear 
magnets, or texturized magnets \cite{Eschrig2015,longrg}. 
Another option 
would be the use of spin-active interfaces in the form of
 magnetic insulators or materials with strong spin-orbit coupling in
 the presence of a Zeeman field \cite{Eschrig2015}. 
 Nonetheless,
 we emphasize 
 that there are
 a number of ways 
 to generate spin-1 triplet 
 correlations that
 would yield essentially the same 
 results 
 presented here.
 Therefore, to simplify the setup and proposed experiment, 
 we consider  the structure sketched in Fig.~\ref{fig1},
with the assumption  that the spin-1 triplet correlations have
been induced in the N region by the Holmium (Ho) layers which can be described by additional terms $(\mathbf{h}\cdot\mathbf{\sigma})\gamma-\gamma(\mathbf{h}\cdot\mathbf{\sigma}^\ast) $, and $\tilde{\gamma}(\mathbf{h}\cdot\mathbf{\sigma})-(\mathbf{h}\cdot\mathbf{\sigma}^\ast) \tilde{\gamma}$ in the Usadel equation Eq.~(\ref{eq:usdl_riccati1}) and (\ref{eq:usdl_riccati2}), respectively. One can also solve the Usadel equation (\ref{eq:usdl}) without the kinetic part and derive new solutions to the bulk superconductors in the presence of an inhomogeneous magnetization,
the effect of which is to 
renormalize $\hat{G}_{\text{S}}$ in Eq.~(\ref{eq:bc}), including the
${\cal C}$ and ${\cal S}$ terms. The Holmium-like magnetization
pattern is coordinate dependent and 
can be described by 
$\mathbf{h}=h_0(\cos\psi,\sin\psi\sin\beta x/a, \sin\psi\cos\beta x/a),$ 
in which $\psi$ and $\beta$ are the apex and azimuthal angles of the cone that constitutes the spiral pattern (see 
Fig.~\ref{fig1}), and  $a$ is the atomic interlayer distance \cite{volvik}. Here we take
the widely used values,
 $\psi=4\pi/9$ and $\beta=\pi/6$ \cite{volvik}.

To determine the signatures of various proximity induced superconducting correlations, we calculate the singlet pair potential $U_{pair}$:
\begin{eqnarray}\label{eq:pairpot}
U_{pair}(\mathbf{R})=-\frac{{\cal N}_0\lambda}{8} \text{Tr}\left\{ \frac{\rho_x-i\rho_y}{2}\tau_z \int d\varepsilon G^K(\varepsilon,\mathbf{R})\right\},~~~
\end{eqnarray}
and the local density of states:
\begin{eqnarray}\label{eq:DOS}
{\cal N}(\varepsilon,\mathbf{R})=\frac{{\cal N}_0}{2}\text{Re}\left[\text{Tr}\{ \hat{G}(\varepsilon,\mathbf{R})\}\right],
\end{eqnarray}
where ${\cal N}_0$ is the density of states per spin at the Fermi level and $\lambda$ is the pairing interaction constant.
\begin{figure}
  \centering
\includegraphics[width=7.50cm,height=7.50cm]{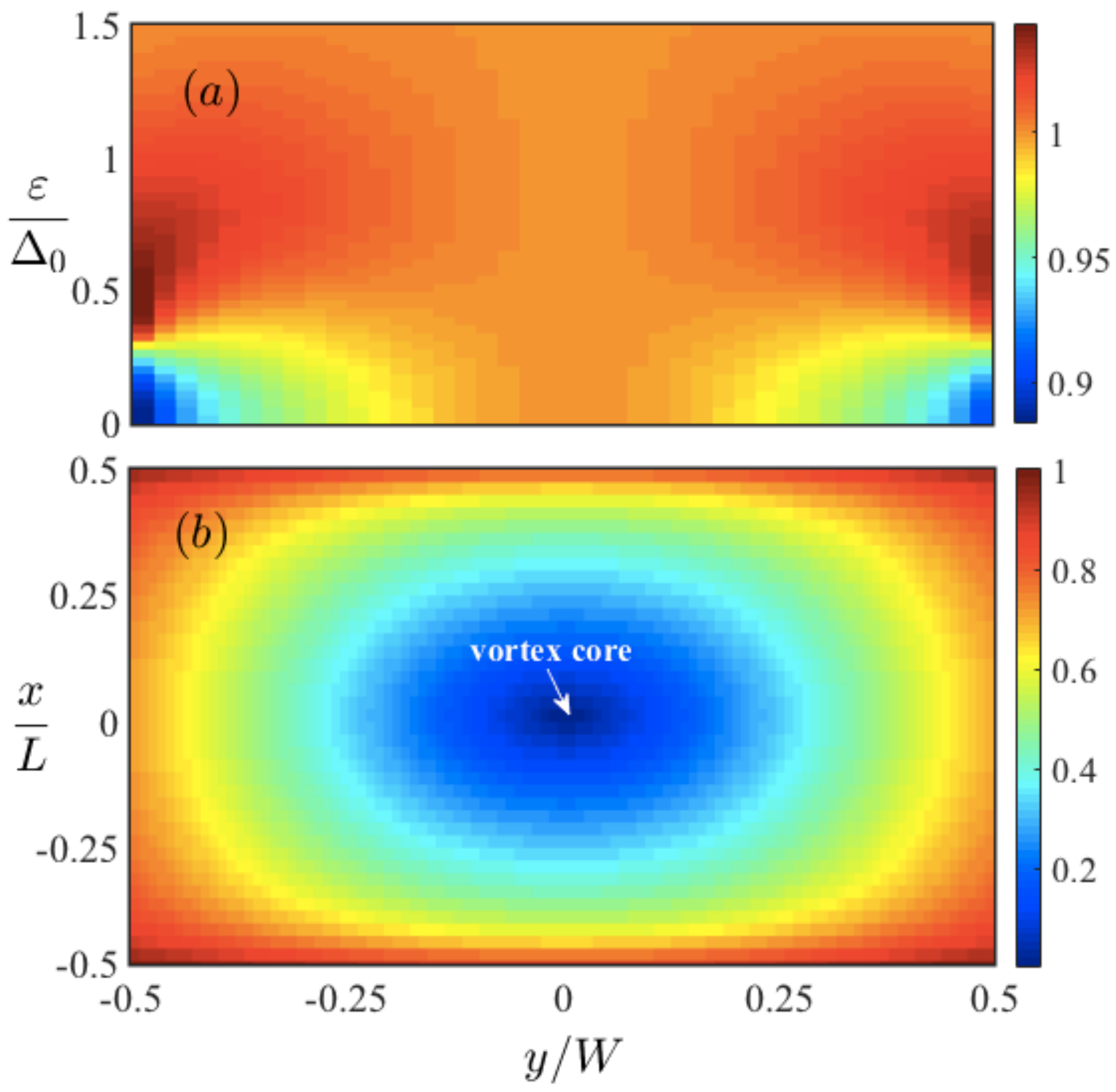}
\caption{\label{fig2} (Color online) ($a$) Color map of normalized local density of states as a function of quasiparticles' energy $\varepsilon$ and location along the junction width $y$ at $x=0$ inside the diffusive normal metal N when the outer Ho layers are off i.e. $h_0=0$. ($b$) Corresponding spatial map of normalized singlet pair potential.}
\end{figure}

\begin{figure*}[t!]
\includegraphics[width=18.0cm,height=4.20cm]{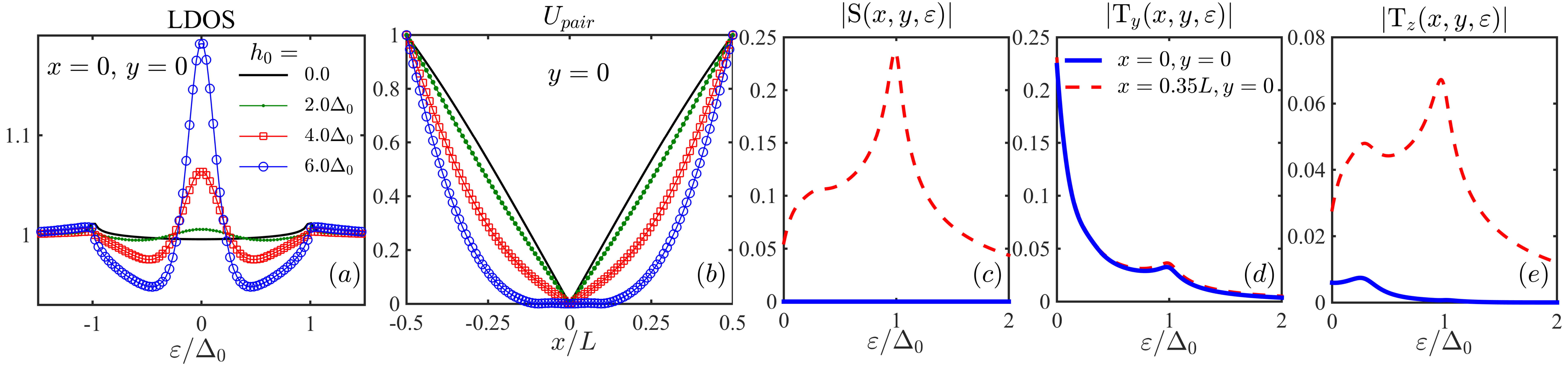}
\caption{\label{fig3} (Color online) ($a$) Local density of states as
  a function of quasiparticles' energy $\varepsilon$ at the vortex
  core, $x=y=0$ (shown in Fig. \ref{fig2}) for four values of internal
  field in the Ho layers, $h_0=0.0, 2.0\Delta, 4.0\Delta,$ and
  $6.0\Delta$. ($b$) Corresponding singlet pair potential, $U_{pair}$,
  along the junction length, $x$, in the same location as the vortex
  core i.e. $y=0$. ($a$)  and ($b$) panels are obtained within the
  \textit{full} proximity limit and the other parameters are identical to those of
  Fig. \ref{fig2}. ($c$)-($e$) show the modulus of singlet
  $|\text{S}(\varepsilon)|$, spin-1 $|\text{T}_y(\varepsilon)|$, and spin-0
  $|\text{T}_z(\varepsilon)|$ triplets against $\varepsilon$ within the \textit{low}
  proximity limit. The solid lines show the correlations at the vortex
core while the dashed lines correspond to a representative location outside of the
vortex core: $x=0.35L, y=0$.}
\end{figure*}

To begin, we present the local DOS and pair potential in Fig.~\ref{fig2} 
for the case $h_0=0$, i.e. the Ho layers in Fig. \ref{fig1} have been replaced
with  normal metals, and consequently
no triplet correlations exist.
To further simplify our analysis, we consider a sufficiently wide junction, $W\gg L$, 
and set the external magnetic field so that only a single magnetic  flux
quantum  $\Phi_0$ passes through the N region. 
We also assume representative values of
$\zeta=4$, the system temperature set at $T=0.05T_c$
(with critical temperature $T_c$), and a superconducting phase difference 
$\phi=\pi$. This choice of $\phi$ only shifts the vortex core to
$x=y=0$ \cite{ZAD_3DTI,longrg} without affecting the final 
outcome. Due to the single magnetic quantum flux in N, 
a single proximity vortex is induced \cite{berg_fr}. 
As seen in panel ($b$), the pair potential vanishes at $x=y=0$,
coinciding with the normal core of the vortex.
To shed more light on the  influence of proximity effects
on the vortex behavior,  
we have also calculated the corresponding LDOS shown in panel ($a$) as a function of the
quasiparticle energy, $\varepsilon$, and location along the junction width, $y$  (at $x=0$). 
It is apparent that the LDOS at $x=y=0$ is equal to unity which clearly demonstrates that no singlet superconducting correlation exist in the vicinity of $x=y=0$, where the singlet pair potential is zero. 
Note that $U_{pair}$ only involves the spin singlet component of 
the
Green function even in the presence of an external magnetic field. 
This can be clearly seen in the low proximity regime where
 the contributions from the singlet and triplet channels can be decomposed \cite{longrg}. 
 Panel ($a$) shows that the LDOS becomes reduced 
 at locations 
 away from $x=y=0$. This can be understood 
 by noting that the 
 {\it singlet} pair correlations are 
 responsible for inducing a minigap in the hybrid 
 structure. 
 This is reflected
in the behavior of the 
 pair potential [panel ($b$)] which shows 
that  $U_{pair}$ increases as one moves away from  the normal core of vortex ($x=y=0$).

Panel ($a$) of Fig.~\ref{fig3} exhibits the LDOS at the center of vortex ($x=y=0$) vs 
the quasiparticle energy $\varepsilon$, and at  differing values of the 
exchange field: $h_0=0, 2\Delta, 4\Delta$, and $6\Delta$. 
Panel ($b$) illustrates the corresponding singlet pair potential along the junction length in the $x$ direction
 (at $y=0$). To have absolute comparisons, the 
 parameters are kept the same as those used in  Fig.~\ref{fig2}.
 The normalized DOS for $h_0=0$ is equal to unity, corresponding
  to the normal phase at the vortex core as discussed in relation to Fig.~\ref{fig2}. 
  Switching $h_0$ to nonzero values immediately induces a peak at zero energy 
  and its amplitude increases with 
  stronger, more inhomogeneous internal fields $\mathbf{h}$. This follows from
   the fact that stronger IFMs can more effectively convert  singlet correlations into  triplet
   ones. 
   Although not shown here, 
   our results
    demonstrated a disappearance of the  ZEP 
   when
    $\mathbf{h}$ is uniform and collinear.
    To clearly determine the type of superconducting correlations responsible for the ZEPs, 
    one can simply consider the singlet pair potential shown in panel ($b$).
     It is apparent that $U_{pair}$ completely vanishes at $x=0$ where the associated LDOS is 
 calculated in panel ($a$). Therefore, the results clearly 
 demonstrate that the only nonvanishing pair correlations at $x=y=0$ are the spin-1 
 triplet pairs, and therefore are responsible for the induction of the
 ZEPs. When the opacity of the interfaces is large enough,
 e.g. $\zeta= 10-20$, the normal and anomalous Green functions can be
 approximated by $|G|\sim 1$ and $|F|\ll 1$ and 
 one can expand the Green function around the bulk solution. This regime
 allows for the
  spin parameterization of the Green function via ${F}({\bm R},\varepsilon)= i
[\text{S}({\bm R},\varepsilon)+{\bm \tau}\cdot{\textbf{T}}({\bm R},\varepsilon)
]\tau_y,$ as exhaustively described in Ref.~\onlinecite{longrg}. The Green
function can then
be decomposed into its singlet $\text{S}({\bm R},\varepsilon)$, spin-1 $\text{T}_y ({\bm R},\varepsilon)$, and
spin-0 $\text{T}_z ({\bm R},\varepsilon)$ triplet components. 
Panels ($c$)-($e$) illustrate
the behaviour of these correlations at the vortex core ($x=y=0$) and
at
$x=0.35L, y=0$, outside of the core. It is clearly seen that the
largest nonvanishing component within low energies at the center of
the
vortex is the odd frequency equal spin component $\text{T}_y ({\bm
  R},\varepsilon)$. Although we have focused on precisely the vortex
core in our calculations, the spin singlet component should be
practically enough suppressed (and triplets dominate)
within a circle with a radius of the magnetic penetration depth around
the vortex core to experimentally reveal the predicted signatures above. 

 It is  known that
vortices in clean superconductors can host  
bound states that are separated in energy by 
an amount $\sim \Delta^2/\varepsilon_F$
\cite{vrx_theo1}. 
These bound states yield a peak in the LDOS of 
a vortex core at the Fermi level \cite{vrx_theo1} that was first 
observed in Ref.~\onlinecite{vrx_bs_exp} 
and then followed up by numerous theoretical and experimental works \cite{vrx_bs_theo,vrx_theo3,vrx_exp1,vrx_exp2,vrx_exp3,vrx4,vrx1,vrx2,vrx3}. 
 These  low-energy states reflect 
relevant details of the bulk gap structure of the superconducting state \cite{balat}.
 It is important to emphasize that the  
 vortices discussed  here are in the diffusive limit 
 where the quasiparticles move in random directions after each collision with
  the scattering sources and $\xi_S$ is much larger than the mean free
  path,  thus excluding  
  the existence of Andreev bound states at the
  vortex cores. 
  To achieve optimal DOS signatures, the STM tip  should be 
  placed near the vortex core, where the odd frequency triplet correlations 
  are revealed through an 
  enhancement of  the zero energy quasiparticle states 
  in  close vicinity of the tip. 
  Finally, it is worth mentioning that
  in light of the specific system parameters used
  for producing spin-1 triplet correlations, 
  the thickness of the
  ${\rm Ho}$ layers and the actual magnetization patterns can play important
  roles in the singlet to triplet  conversion process
  \cite{longrg,crnt_1,crnt_2}. In this work, the two ${\rm Ho}$ layers are
  considered identical and of thickness 10nm \cite{longrg,crnt_1,crnt_2}.

{\it Conclusions.}~
To summarize, 
motivated by recent experimental progress 
related to the proximity induced vortices \cite{vrtx_exp}, we have
  proposed an experimentally accessible platform 
  that utilizes the cores of proximity vortices  
  to isolate the equal spin triplet pairings \cite{Buzdin2005,first}. 
  We showed that a 
  proximity-induced vortex can be generated in the normal layer of a 
  two dimensional diffusive S-Ho/N/Ho-S junction by  applying 
  an external magnetic field to the junction plane,  with the Holmium (Ho) layers 
  serving as  sources of spin-1 triplet correlations.
  We then demonstrated that one can directly probe the equal spin triplet pairings 
  via a tunneling spectroscopy experiment at the normal core of the vortex. 

\acknowledgements
M.A. would like to thank G. Sewell for his valuable instructions in the numerical
parts of this work. A.Z. was financially supported by the Swedish
Research Council Grant No. 642-2013-7837. K.H. is supported in part by ONR and a grant
of HPC resources from the DOD HPCMP.

\end{document}